\documentclass[conference,a4paper]{IEEEtran}
\IEEEoverridecommandlockouts
\usepackage{cite}
\usepackage{amsmath,amssymb,amsfonts}
\usepackage{algorithmic}
\usepackage{graphicx}
\usepackage{caption}
\usepackage{subcaption}
\usepackage{textcomp}
\usepackage{xcolor}
\usepackage[left=1.35cm,right=1.35cm,top=1.9cm,bottom=4.1cm]{geometry}

\usepackage{bbm}
\def\BibTeX{{\rm B\kern-.05em{\sc i\kern-.025em b}\kern-.08em
    T\kern-.1667em\lower.7ex\hbox{E}\kern-.125emX}}
\usepackage[subtle,tracking=normal]{savetrees}
\begin{document}

\title{Graph Neural Networks for Power Allocation in Wireless Networks with Full Duplex Nodes\\

}

\newcommand\blfootnote[1]{%
  \begingroup
  \renewcommand\thefootnote{}\footnote{#1}%
  \addtocounter{footnote}{-1}%
  \endgroup
}

\author{\IEEEauthorblockN{Lili Chen, Jingge Zhu and Jamie Evans}
\IEEEauthorblockA{Department of Electrical and Electronic Engineering, University of Melbourne, Australia \\
Email: lilic@student.unimelb.edu.au, jingge.zhu@unimelb.edu.au, jse@unimelb.edu.au\\
}

}

\maketitle

\begin{abstract}
Due to mutual interference between users, power allocation problems in wireless networks are often non-convex and computationally challenging. Graph neural networks (GNNs) have recently emerged as a promising approach to tackling these problems and an approach that exploits the underlying topology of wireless networks. In this paper, we propose a novel graph representation method for wireless networks that include full-duplex (FD) nodes. We then design a corresponding FD Graph Neural Network (F-GNN) with the aim of allocating transmit powers to maximise the network throughput. Our results show that our F-GNN achieves state-of-art performance with significantly less computation time. Besides, F-GNN offers an excellent trade-off between performance and complexity compared to classical approaches. We further refine this trade-off by introducing a distance-based threshold for inclusion or exclusion of edges in the network. We show that an appropriately chosen threshold reduces required training time by roughly $20\%$ with a relatively minor loss in performance.

\end{abstract}


\section{Introduction}\label{sec:introduction}


Power allocation is crucial to the performance of wireless communications networks, especially under time-varying channel conditions. However, power allocation is often a non-convex problem due to the interference between channels.
For the sum rate maximisation problem, a number of classical approaches have been proposed in the literature (see for example \cite{naderializadeh2014itlinq, shi2011iteratively}). Nevertheless, they are computationally intensive in large wireless networks and thus inappropriate for practical implementation \cite{sun2018learning}.

More recently, there has been significant interest in deep learning-based approaches for solving power allocation problems.
For example,  multi-layer perceptrons (MLPs), which were inherited from image identification tasks, are now widely used for power allocation in wireless communication \cite{sun2018learning}.
However, as the network size increases, the performance of these algorithms degrades dramatically \cite{shen2020graph}. On the one hand, it is computationally expensive to train MLPs with high-dimensional data. On the other hand, this architecture may fail to exploit the graph structure of wireless networks.

\blfootnote{The work was supported by the Melbourne Research Scholarship of the University of Melbourne and in part by Australian Research Council under project DE210101497.}
To remedy these drawbacks, several researchers have applied graph neural networks (GNNs) to the power allocation problem due to their ability to exploit the network structure. Besides, some researchers also prove that GNNs may perform better than MLPs when it comes to graph-structured data \cite{shen2021neural}. Message-passing graph neural networks (MPGNNs) are proposed in \cite{shen2020graph} to find the optimal power allocation with unsupervised learning. Comprehensive simulations demonstrate that the MPGNNs have a similar performance to the weighted sum MSE minimization (WMMSE) algorithm \cite{shi2011iteratively} with less computational complexity and the model has good generalisation capacities. Heterogeneous GNNs with a novel graph representation are proposed in \cite{zhang2021scalable} and \cite{li2022heterogeneous} to allocate power in device-to-device (D2D) and cell-free massive Multiple-Input-Multiple-Output (MIMO) networks, respectively.


In most cases, the power allocation problem is formulated in the context of half-duplex (HD) transmission. However, as a promising technique to provide the potential of doubling the capacity compared to conventional HD transmission \cite{sabharwal2014band}, full-duplex (FD) transmission has drawn much attention recently. 
To the best of our knowledge, current GNN-based methods in power allocation only focus on HD transmission, and only a few papers focus on FD transmission such as \cite{li2022heterogeneous}. However, the authors preprocess the node feature to guarantee the scalability of GNN, which introduces additional non-negligible computational time. To address this issue, we propose a novel graph representation method for FD transmission in wireless networks. The node feature can be directly fed into GNN without loss of scalability. Then, we devise the corresponding full-duplex Graph Neural Network (F-GNN) to optimising the power allocation in D2D networks with FD nodes.
We conduct extensive experiments to evaluate the effectiveness of the proposed graph representation and F-GNN. 
Simulation results show that F-GNN slightly outperforms the advanced optimisation-based WMMSE algorithm with much lower time complexity.

Intuitively, unpaired transmitters and receivers in D2D networks are more likely to be far apart. Since the interference will decay with distance, we expect that removing edges between nodes that are distant, should further reduce the computational complexity without impacting the performance significantly. In \cite{shen2020graph}, the authors assume the channel state is used in GNNs when the distance is within a specific threshold to reduce the training overload.  Similarly, the authors in \cite{wang2022learning} apply the distance-based threshold to alleviate the negligible interference between unpaired transmitters and receivers. However, the threshold is usually randomly chosen without further justification. In this paper, we investigate the impact of distance-based thresholds on both performance and time complexity. In particular, we derive the explicit relationship between the threshold and the expected time complexity.


\section{Preliminaries}

\subsection{System Model}\label{sub:systemmodel}

We consider the power allocation problem in a single-hop D2D communication network with $K$ users,  where multiple single-antenna devices share the same spectrum. Here, we assume the first $T_1$ devices can use the wireless FD transmission with different frequency bands, while the other devices are on HD transmission. We denote the index set for FD communication devices by $\mathcal{T}_1 = \{1,2,...,T_1\}$ and the index set for HD communication devices as transmitter by $\mathcal{T}_2 = \{T_1+1,T_1+2,...,T_1+T_2\}$. For each $k$ in $\mathcal{T}_1$ or $\mathcal{T}_2$, we define $I(k)$ to be the index of the intended receiver. The received signal at $I(k)$-th receiver for any $k \in \mathcal{T}_1$ with FD transmission is given by
\begin{equation}
y_{I(k)}=h_{k, I(k)} s_{k}+\sum_{j\in  \mathcal{T}_1 \cup \mathcal{T}_2  \backslash\{k\}} h_{ j,I(k)} s_{j}+ n_{I(k)} + z_{I(k)}, k \in \mathcal{T}_1,
\end{equation}
\noindent where $h_{k,I(k)} \in \mathbb{C}$ represents the communication channel between $k$-th transmitter and its intended $I(k)$-th receiver, $h_{ j,I(k)} \in \mathbb{C}$ represents the interference channel between $j$-th transmitter and $I(k)$-th receiver. We denote $s_{k} \in \mathbb{C}$ as the data symbol for the $k$-th transmitter, $n_{I(k)} \sim \mathcal{C} \mathcal{N}\left(0, \sigma_{I(k)}^{2}\right)$ and $z_{I(k)} \sim \mathcal{C} \mathcal{N}\left(0, \gamma^2_{I(k)} \right)$ as the additive Gaussian noise and self-interference for the $I(k)$-th receiver, respectively. The received signal at $I(k)$-th receiver for any $k \in \mathcal{T}_2$ with HD transmission is given by
\begin{equation}
y_{I(k)}=h_{k,I(k)} s_{k}+\sum_{j\in  \mathcal{T}_1 \cup \mathcal{T}_2 \backslash\{k\} } h_{ j,I(k)} s_{j} +n_{I(k)}, k \in \mathcal{T}_2,
\end{equation}
here, we assume each transmitter either in FD or HD modes only has one intended receiver. The signal-to-interference-plus-noise ratio (SINR) for the $I(k)$-th receiver with FD transmission is given by
\begin{equation}
\operatorname{SINR}_{I(k)}=\frac{\left|h_{k,I(k)}\right|^{2} p_{k}}{ \sum_{j\in  \mathcal{T}_1 \cup \mathcal{T}_2 \backslash\{k\}} \left|h_{j, I(k)}\right|^{2} p_{j}  +\sigma_{I(k)}^{2} + \gamma^2_{I(k)}},  k \in \mathcal{T}_1,
\end{equation}
where $p_{k}=\mathbb{E}\left[s_{k}^{2}\right]$ is the power of the $k$-th transmitter, and we have the constraints $0 \leq p_{k} \leq P_{max}$,
where $P_{\max}$ is the maximum power constraint for transmitters.
The SINR for $I(k)$-th receiver with HD transmission is given by
\begin{equation}
\operatorname{SINR}_{I(k)}=\frac{\left|h_{k,I(k)}\right|^{2} p_{k}}{ \sum_{j\in  \mathcal{T}_1 \cup \mathcal{T}_2 \backslash\{k\} } \left|h_{j, I(k)}\right|^{2} p_{j}+ \sigma_{I(k)}^{2}},  k \in \mathcal{T}_2,
\end{equation}
We denote $\mathbf{p}=\left[p_{1}, \cdots,    p_{K}\right]$ as the power allocation vector. For a given power allocation vector $\mathbf{p}$ and channel information $\left\{h_{i j}\right\}_{i \in \mathcal{T}_1 \cup \mathcal{T}_2, j \in I(i)}$, the achievable rate $\mathcal{R}_{I(k)}$ of the $I(k)$-th receiver is given by
\begin{equation}
\mathcal{R}_{I(k)}(\mathbf{p})=\log_2 \left(1+\operatorname{SINR}_{I(k)} \right), \quad k \in \mathcal{T}_1 \cup \mathcal{T}_2.
\end{equation}

We assume that the channel coefficients are fixed in each time slot, thus the output for the power allocation algorithm is $\mathbf{p}$ which is based on the channel information. The objective is to maximise the performance under maximum power constraints, which is formulated as,
 \begin{equation}
    \begin{array}{cl}
    \underset{\mathbf{p}}{\operatorname{maximise}} & \sum_{k} w_{k}  \mathcal{R}_{I(k)}(\mathbf{p}), \\
    \text { subject to } & 0 \leq p_{k} \leq P_{\max }, \forall k \in \mathcal{T}_1 \cup \mathcal{T}_2,
    \end{array}
    \label{eq:weightsumrate}
\end{equation}   
where $w_{k}$ is the weight for the ${k}$-th transmitter.

\subsection{Self-interference Model}
It is impossible to completely eliminate self-interference in practical situations, despite the advanced self-interference mitigation techniques applied. Therefore, we assume the self-interference is zero-mean Gaussian distribution and model the variance of it caused by imperfect mitigation as\cite{rodriguez2014performance},
\begin{equation}
\gamma^2=\eta p^{\lambda},
\label{eq:selfinterference}
\end{equation}
where $p$ is the transmitted power in FD transmission, $\eta$ and  $\lambda \in [0,1] $ are parameters that indicate the attribute of the mitigation technique \cite{duarte2012experiment}. The smaller $\eta$ and $\lambda$ are the more superior self-interference mitigation is. In this paper, we consider three self-interference parameters $\lambda \in \{0.1, 0.5, 1\}$ which represent different levels of mitigation. 


\section{Graph-Based Neural Network design for  Power Allocation}

In this section, we propose a general framework based on a novel graph representation for wireless networks with full duplex nodes to maximise the target function in \eqref{eq:weightsumrate}.


\subsection{Graph Representation}
\begin{figure}[htbp]
\centerline{\includegraphics[width=8cm]{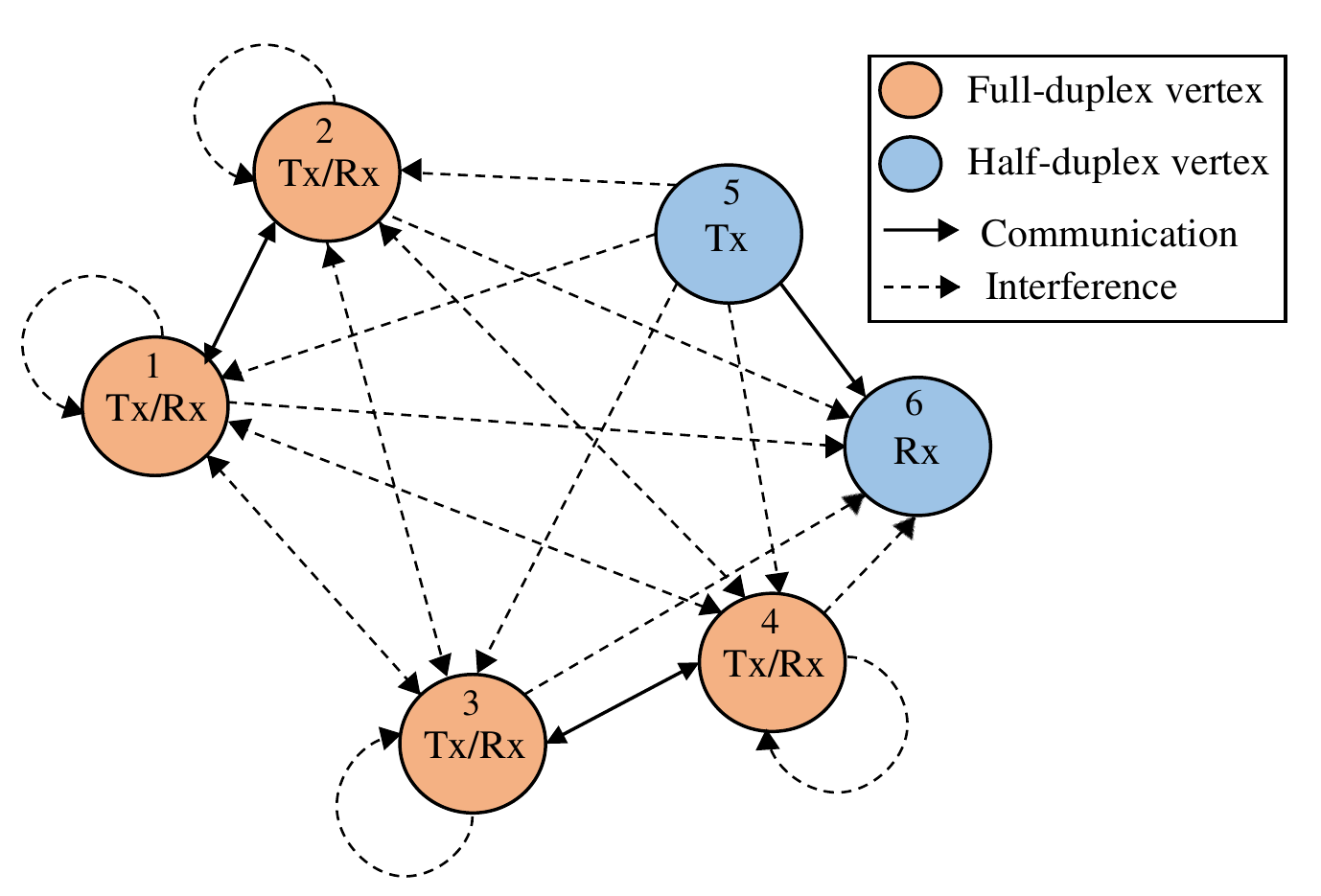}}
\caption{D2D Communication Network with full-duplex transmission when $K=6$.}
\label{fig:D2Dfullduplex}
\end{figure}

In our problem, we can model transmitters and receivers in wireless networks as vertices and the interference between them as edges, see Fig.~\ref{fig:D2Dfullduplex} for example. Since there exists self-interference between transmitters and receivers within the FD transmission, we add a self-loop to these vertices to model such an effect. In the proposed framework, the vertices can possibly perform in HD or FD transmissions. In order to distinguish the working modes for vertices, we propose a novel graph representation method as follows. In the first step, we adopt a similar idea as in \cite{li2016binary} by duplicating the vertices working in the FD transmission, and we treat one vertex as the transmitter and the other as the receiver. By doing so, all vertices are converting to HD transmission mode equivalently. Then, we can model the self-interference channel as edges between these two vertices. 

\begin{figure}[htbp]
\centerline{\includegraphics[width=8.5cm]{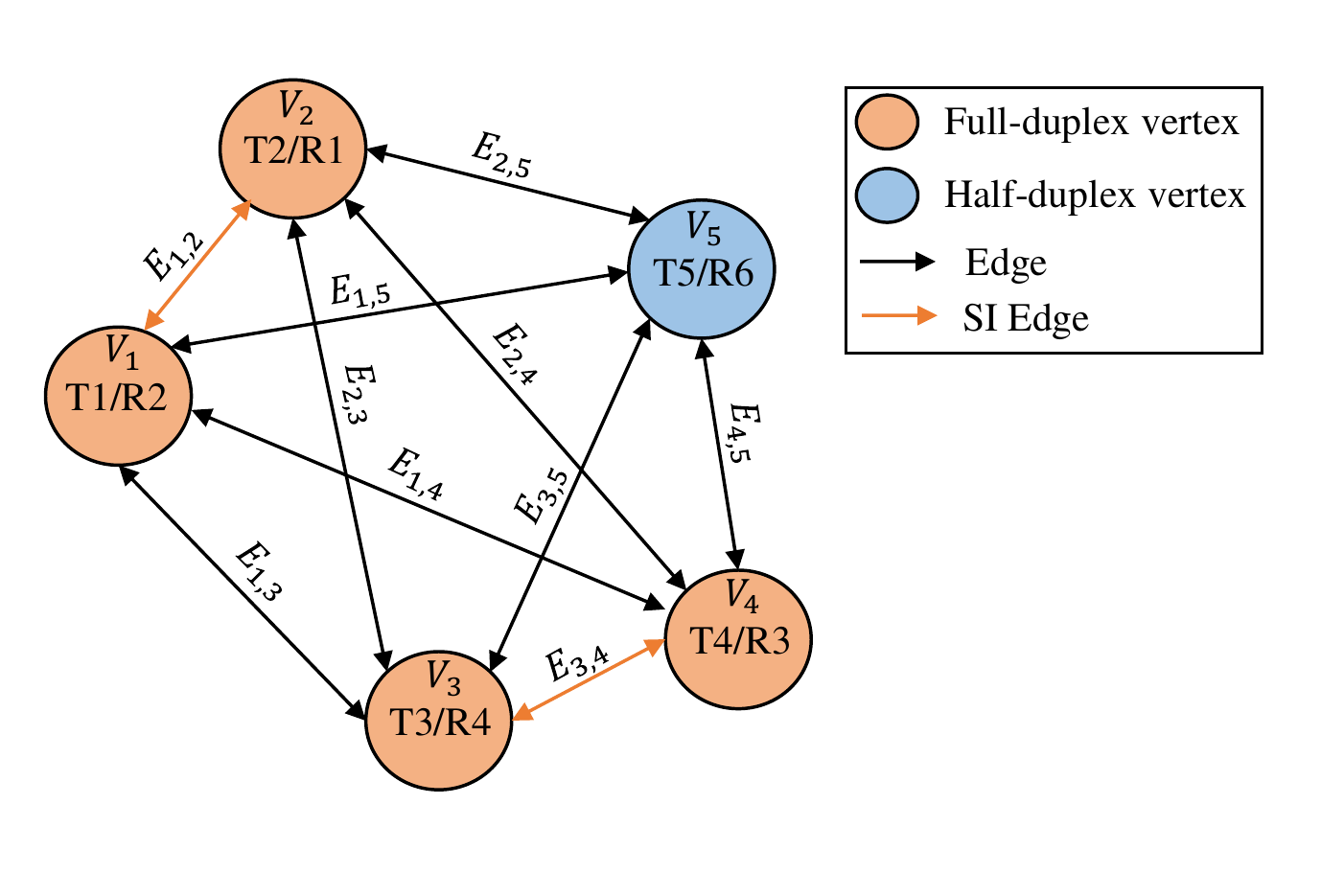}}
\caption{Graphical model of Fig.~\ref{fig:D2Dfullduplex}}
\label{fig:D2Dfullduplexgraph}
\end{figure}

In the weighted sum-rate maximisation problem, the channels between transmitters and receivers can be generally categorised into communication and interference channels. Since these two quantities play different roles in our objective function \eqref{eq:weightsumrate}, it is better to distinguish them in graph modelling. To this end, in the second step, we construct a new graph as shown in Fig.~\ref{fig:D2Dfullduplexgraph} by aggregating the transmitter and receiver pair as a vertex, the self-interference channel and the interference channels between two vertices as a SI edge and an edge, respectively. The quantities such as communication channel information, direct distance, self-interference model information and weights can be regarded as the vertex feature while the interference channel information and interfering distance can be regarded as the edge feature.

Let $\mathcal{V}$ and $\mathcal{E}$ denote the set of vertices and edges of a graph $G$, respectively. The set of the neighbours of $v \in \mathcal{V}$ is defined as $\mathcal{N}(v)$.
Let $V_v$ and $E_{v,u}$ represent vertex features of vertex $v$ and edge features between vertex $v$ and vertex $u$, respectively. 
With definitions in place, we define the vertex features of the vertices to be
\begin{equation}
V_{v}= \begin{cases}h_{ v,I(v)} , w_{v}, d_{ S} & \text { if } v\in \mathcal{T}_1, \\ h_{ v,I(v)} , w_{v}, d_{ v,I(v)}, & \text { if } v\in \mathcal{T}_2. \end{cases}
\end{equation}
where $h_{v,u} \in \mathbb{C}$ and $d_{v,u} \in \mathbb{R}$ are the channel coefficient and distance between $v$-th transmitter and $u$-th receiver, $w_v$ is the weight for $v$-th transmitter, and $d_{S} \in \mathbb{R}$ is the distance between interfering antennas in full-duplex nodes. 
We define the edge features to be
\begin{equation}
E_{v,u}= \begin{cases}z_{ v} , z_{ u}, d_{ S},d_{ S}, & \text { if } v ,u  \in \mathcal{T}_1, u = I(v), \\ h_{ u,I(v)} , h_{v,I(u)}, d_{ u,I(v)},d_{ v,I(u)} & \text { otherwise. }  \end{cases}
\end{equation}
where 
$z_v \in \mathbb{C}$ is the self-interference for $v$-th receiver.

\subsection{Graph Neural Networks}
GNNs are firstly proposed to process graph-based data since they can exploit the graph structure of data to extract useful information. Specifically, the vertex updates its embedding feature vector by using the information from its previous layer and the aggregated information from its neighbour. Mathematically, the update rule of the $l$-th layer at vertex $v$ in GNNs is given as follows \cite{xu2018powerful}:
\begin{equation}
\begin{aligned}
&\alpha_{v}^{(l)}=\text { AGGREGATE }^{(l)}\left(\left\{m_{u}^{(l-1)}: u \in \mathcal{N}(v)\right\}\right), \\
&m_{v}^{(l)}=\text { COMBINE }^{(l)}\left(m_{v}^{(l-1)}, \alpha_{v}^{(l)}\right),
\end{aligned}
\end{equation}
where $\alpha_{v}^{(l)}$ is the aggregated feature vector by vertex $v$ from its neighbours and $m_{v}^{(l)}$ is the embedding feature vector of vertex $v$ at the $l$-th layer. $\text { AGGREGATE }^{(l)}$ and $\text { COMBINE }^{(l)}$ are two functions defined by the user of the GNN.
 
\begin{figure}[htbp]
\centerline{\includegraphics[width = 5.5cm]{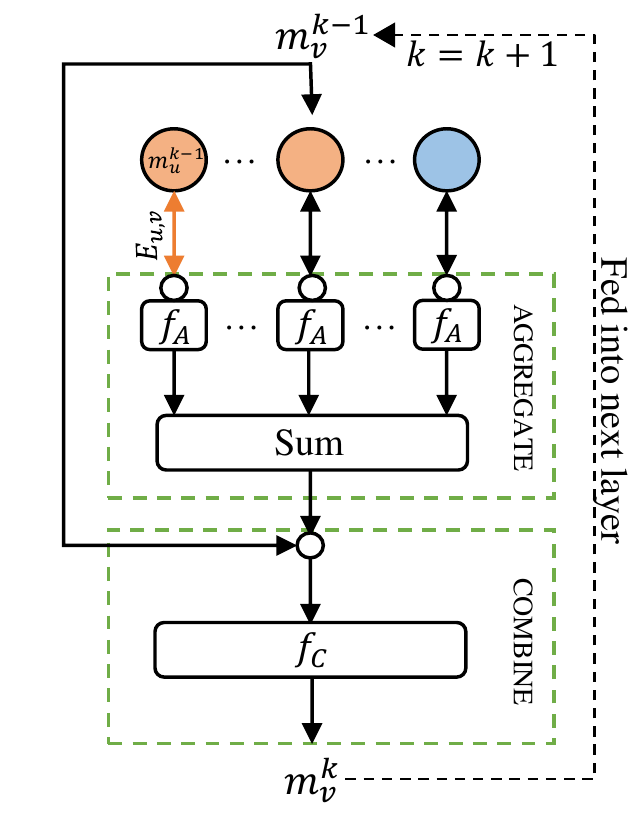}}
\caption{The structure of the proposed F-GNN}
\label{fig:GNNdtructure}
\end{figure}

Owing to the universal approximation capability of MLP \cite{hornik1989multilayer}, we will adopt MLP for both aggregating information from a local graph-structured neighbourhood and combining its own features with the aggregated information. 
Besides, the aggregation step is expected to retain the permutation invariance property, i.e., the aggregated information is invariant for different input orders of neighbour vertices. We can achieve this by using a permutation-invariant function after the MLP, such as sum and max operations, to combine a set of aggregated information from its neighbour into a single set. Here, we choose sum as it can preserve all the information.
The updating rule of the proposed GNN at $l$-th layer is given by
\begin{equation}
\begin{aligned}
& \alpha_{v}^{(l)}=\operatorname{SUM}\left(\left\{ f_A \left(m_u^{(l-1)}, E_{vu} \right), \forall u \in \mathcal{N}(v)\right\} \right),  \\
& p_{v}^{(l)} =f_C \left(\alpha_{v}^{(l)}, m_{v}^{(l-1)}\right),
\end{aligned}
\label{eq:gnn}
\end{equation}
where $\alpha_{v}^{(l)}$ represents the aggregated information by vertex $v$ from its neighbours, $m_v^{(l)} = \{V_v, p_v^{(l)}\}$ represents the embedding feature vector of vertex $v$, $p_v^{(l)}$ represent the allocated power for vertex $v$ and $f_{A}$, $f_{C}$ are two 3-layer fully connected neural networks with hidden sizes $\{8,16,32\}$ and $\{36,16,1\}$, respectively. Here, we initialise the power $p_{v}^{(0)} = P_{\max}$. An illustration of the F-GNN structure is shown in Fig.~\ref{fig:GNNdtructure}.
\section{Experiments and Result}\label{sec:experiments}
In this section, we provide simulation results for three types of self-interference models to demonstrate the benefits of the proposed F-GNN architecture. 
\subsection{Simulation Setup}
We consider a channel with large-scale fading and Rayleigh fading as in \cite{liang2019towards}. For the system setup, the channel state information (CSI) is formulated as, $h_{ v,u} = \sqrt{\frac{1}{1+d_{v,u}^{2}}} r_{v,u}$,
\noindent where $r_{v,u} \sim \mathcal{C}\mathcal{N}\left(0, 1\right)$ and $d_{v,u}$ is the distance between $v$-th transmitter and $u$-th receiver. Here, we consider $K$ users within a $100 \times 100$ $m^2$ area. The transmitters are placed in this area randomly while each receiver is placed randomly within 2$m$ to 10$m$ away from the corresponding transmitter. Due to space limitation, this paper mainly focuses on the case when the total number of users with FD transmission $T_1 = 0.5K$. However, our proposed algorithm can also be generalised to different ratios. Since $\eta$ in \eqref{eq:selfinterference} has negligible influence on the normalised performance in our experiments, we set $\eta$ to 0.01 and the distance between interfering antennas in an FD vertex as 40$cm$ \cite{duarte2012experiment}. 
 To achieve a good power allocation strategy, we choose the negative sum rate to be the objective loss function, which can be minimised by the back-propagation algorithm \cite{liang2019towards}. In particular, the loss function is expressed in \eqref{eq:lossfunction}, where $\theta$ represents the learnable parameters of the GNN, and $p_{i}(\theta)$ denotes the power allocation generated by F-GNN. $\boldsymbol{H}=\left[\boldsymbol{h}_{1}, \cdots, \boldsymbol{h}_{K}\right]^{T}$ is the channel matrix, where $\boldsymbol{h}_{i}=$ $\left[h_{i1 }, \cdots, h_{i K}\right], i=1, \cdots, K .$ We use $\hat{\mathbb{E}}$ to denote the expectation with respect to the empirical distribution of the channel samples. In practice, we generate 10000 training samples for calculating the empirical loss and 1000 testing samples for evaluation. We assumed that full CSI is available to the algorithms. However, our proposed algorithm only needs partial CSI (a subset of full CSI) to achieve reasonably good performance as will be discussed in Section~\ref{sec:threhsold}.
 We use Adam optimiser \cite{kingma2014adam} with a learning rate of 0.005 in training.
\begin{figure*}[htp]
\begin{equation}
\begin{aligned}
L(\theta)= & -\hat{\mathbb{E}}_{\boldsymbol{H}}\Bigg\{\sum_{k \in \mathcal{T}_1} w_{k} \log _{2}\left(1+\frac{\left|h_{k,I(k)}\right|^{2} p_{k}(\theta)}{ \sum_{j\in  \mathcal{T}_1 \cup \mathcal{T}_2 \backslash\{k\}} \left|h_{j, I(k)}\right|^{2} p_{j}(\theta)  +\sigma_{I(k)}^{2} + I_{I(k)}}\right) \\
& + \sum_{k \in \mathcal{T}_2} w_{k} \log _{2}\left(1+\frac{\left|h_{k,I(k)}\right|^{2} p_{k}(\theta)}{ \sum_{j\in  \mathcal{T}_1 \cup \mathcal{T}_2 \backslash\{k\} } \left|h_{j, I(k)}\right|^{2} p_{j}(\theta)+ \sigma_{I(k)}^{2}}\right) \Bigg\} \\
\end{aligned}
\label{eq:lossfunction}
\end{equation}
\par\noindent\rule{\textwidth}{1pt}
\end{figure*}
To validate the effectiveness of F-GNN, we compare it with the following two algorithms:
\begin{itemize}
    \item WMMSE \cite{shi2011iteratively}: This is the advanced optimisation-based algorithm for power allocation in wireless networks, also see  \cite{sun2018learning, liang2019towards, shen2020graph} for references.
    \item Baseline: We allocate the maximum power $P_{\text {max}}$ for $L$ pairs which have the $L$th-largest communication channel gain among all pairs, while the rest are set as 0. Here we consider $L = 0.5K$. 
\end{itemize}
\subsection{Performance Comparison}
First, we set $w_{k} = 1$ in \eqref{eq:weightsumrate}, the object function becomes the typical sum rate maximisation problem. The performance result under different self-interference models is shown in Table~\ref{Tab:sumratemaximization}. From this, we observe that the proposed F-GNN achieves similar performance under different $\lambda$, which indicates that the model is robust to different levels of self-interference. This observation also pertains even if we have a larger number of users.

\begin{figure}[htbp]
\centerline{\includegraphics[width = 7.5cm]{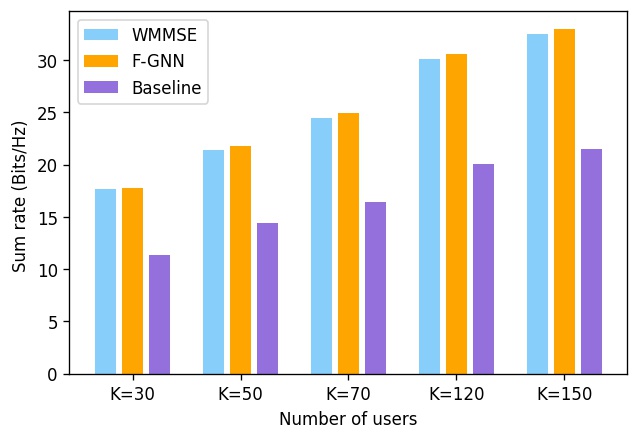}}
\caption{Average weighted sum rate under the full-duplex network. }
\label{fig:weighsumratepathlossmode}
\end{figure}

For simplicity, we only consider the case with $\lambda = 0.5$ for the rest of the experiments unless specified. Now, we let $w_k$ be randomly drawn from $\mathcal{U} (0,1)$ and the experimental result is shown in Fig.~\ref{fig:weighsumratepathlossmode}. 
It can be seen that F-GNN can achieve similar performance to WMMSE. We can also observe that the performance gap increases as $K$ increases. This is possibly because F-GNN can extract the underlying graph structure better and thus will be more useful when dealing with complicated networks. Besides, this will also induce good generalisation capacities which can be seen in section~\ref{sec:Generalization}.
\begin{table}[htbp]
\caption{Average sum rate that is normalised by the performance from WMMSE under different self-interference models.}
\centering
\begin{tabular}{|l|c|c|c|c|}
\hline & $\lambda = 0.1 $ & $\lambda = 0.5$ & $\lambda = 1 $  \\
\hline $K=30$   & $ 100.4\%$ & $ 100.1\% $ & $ 100.7\% $    \\
\hline $K=50$   & $ 100.8\%$ & $100.8 \% $ & $ 100.7\% $    \\
\hline $K=70$   & $ 100.7\%$ & $100.8 \% $ & $ 100.9\% $    \\
\hline $K=120$   & $ 100.9\%$ & $ 100.9\% $ & $ 101.1\% $    \\
\hline $K=150$   & $ 101.1\%$ & $ 101.2\% $ & $ 101.2\% $    \\
\hline
\end{tabular}
\label{Tab:sumratemaximization}
\end{table}
\subsection{Generalisation}\label{sec:Generalization}
Apart from the ability to achieve similar performance to WMMSE, another major advantage of F-GNN is the generalisation capability.
We first train the F-GNN with $K=50$, then we set the number of users in the test set to $\{70, 100, 120, 150\}$ while the area remains the same. The generalisation results are shown in Table \ref{Tab:Generalizationpathloss}. Overall, the proposed model exhibits good generalisation capacities in terms of different users number. Even when the density is three times larger than the training set, F-GNN still slightly outperforms WMMSE. 
\begin{table}[htbp]
\centering
\caption{Average sum rate under the full-duplex network. The sum rate is normalised by the performance from WMMSE for each $K$.}
\begin{tabular}{|l|c|c|c|c|c|}
\hline & $K=70$ & $K=100 $ & $K=120 $ & $K=150 $  \\
\hline GNN   & $ 100.8 \%$ & $ 100.8\% $ & $ 100.7\% $ & $ 100.6\% $   \\
\hline
\end{tabular}
\label{Tab:Generalizationpathloss}
\end{table}
\subsection{Time Complexity}\label{subsection:time_complexity}
From the perspective of the time complexity, in each layer, each edge passes through the neural network $f_A$ and sum operation while each vertex goes through the neural network $f_C$. Thus, the time complexity for GNN in each layer is $\mathcal{O}(|\mathcal{V}|+|\mathcal{E}|)$\cite{shen2020graph}. 
The average testing times for WMMSE and F-GNN under the same experimental settings are shown in Table~\ref{tab:fullduplex_time}. We observed that F-GNN is notably faster than WMMSE. As $K$ increases, F-GNN becomes more efficient compared to WMMSE. 
\begin{table}[htbp]
\centering
\caption{Average testing time in milliseconds for the F-GNN under different settings.}
\begin{tabular}{|c|c|c|c|c|}
\hline & $K=50$ & $K=100$ & $K=500$ & $K=1000$\\
\hline GNN & $ 0.92 $  & $1.36 $  & $ 10.39$ & 41.05  \\
\hline WMMSE & $ 65.63$ & $ 260.13$ & $ 5997.23$ & 24430.73 \\
\hline
\end{tabular}
\label{tab:fullduplex_time}
\end{table}

\section{Threshold} \label{sec:threhsold}

In Section~\ref{sec:experiments}, we assume the networks are fully connected. However, as the interference will decay as the distance increases, we conduct similar experiments but exclude edges whose distances are greater than a specific threshold $t$ meters and investigate the impact of the threshold on both sum-rate performance and time complexity.
To provide a performance upper bound, we conduct the WMMSE with the untruncated networks.

\subsection{Performance and Training Time}
We apply the same threshold to both the training and test set for edge truncation. Here we generate 14 different training and test sets, each with different $t$ ranging from 10 to 140 $m$. The performance of different thresholds under $K=50$ is shown in Table~\ref{tab:threshold_performance}. When the threshold increases, fewer edges will be truncated and the graph will maintain more information from the original graph. Training with a graph closer to the original one will likely have better performance but will have longer training time as more edges are included. Hence, there is a trade-off between performance and time complexity. We also noticed that both the performance and time complexity become stable after $t=100$$m$. We found out that around $3\%$ of the total edges are within $t=100$ and $t=140$$m$ through experiments and it resulted in a negligible effect. Therefore, We only consider $t$ between 10 and 100$m$ in the next section.
\begin{table*}[htbp]
\centering
\caption{Normalised performance and training time versus threshold.}
\begin{tabular}{|c|c|c|c|c|c|c|c|c|c|c|c|c|c|c|}
\hline t & $10$ & $20$ & $30$ & $40$& $50$ & $60$ & $70$ & $80$& $90$ & $100$ & $110$ & $120$& $130$ & $140$\\
\hline GNN($\%$) & $91.05$  & $96.0 $  & $97.9 $ & $99.3$ & $99.6$  & $99.8$  & $100.1 $ & $100.3$  & $100.5 $ & $100.6$ & $100.7 $ & $100.7$  & $100.7 $ & $100.8$\\
\hline Time($s$) & $510 $ & $544 $ & $564$ & $583 $ & $603  $  & $620 $  & $630 $ & $641$ & $650 $ & $658 $ & $660 $ & $662 $  & $664$ & $665$\\
\hline
\end{tabular}
\label{tab:threshold_performance}
\end{table*}
\begin{figure*}[h]
\begin{equation}
\begin{aligned}
F(z)= \begin{cases}-\frac{8}{3 a^{3}} z^{\frac{3}{2}}+\frac{\pi z}{a^{2}}+\frac{z^{2}}{2 a^{4}} & \text { if } 0<z \leqslant a^{2}  \\ \frac{1}{3}
-\frac{2+\pi}{a^{2}} z+\frac{4}{a^2} (\arcsin \left(\frac{a}{\sqrt{z}}\right) z + a \sqrt{z-a^2})+\frac{8}{ 3a^{3}} ({z-a^{2}})^\frac{3}{2} -\frac{z^2}{2 a^{4}}  & \text { if } a^{2}<z \leqslant 2 a^{2}  \end{cases}
\end{aligned}
\label{eq:thresholdedges}
\end{equation}
\par\noindent\rule{\textwidth}{1pt}
\end{figure*}
\subsection{Performance Comparison}

\begin{figure}[h]
	\centering
	\includegraphics[width = 7.8cm]{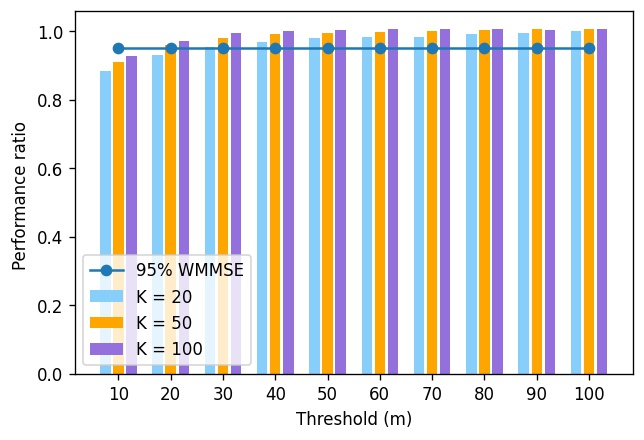}
	\caption{Performance versus threshold }
	\label{Fig:threshold_Generalization_performance}
\end{figure}

We further investigate the effect of the threshold by applying it to different $K$. The performance comparisons are shown in Fig.~\ref{Fig:threshold_Generalization_performance}, where the results for each $K$ are normalised over the sum rate achieved by WMMSE.

Overall, our model can achieve reasonably high performance (e.g, 95$\%$ of the sum rate achieved by WMMSE) when the threshold is greater than 30. Given a specific $K$, we define the "ideal threshold" to be the smallest threshold that can achieve 95$\%$ performance of WMMSE. The ideal threshold for different $K$ is shown in Table~\ref{tab:optimal_threshold}. We also observe that when $K$ increases, the ideal threshold decreases. This is because the graph network becomes denser as the number of users increases, and thus F-GNN has enough interference to learn even with a lower threshold. From Table~\ref{tab:threshold_performance}, we observe that the training time increases as the threshold increases. Therefore, selecting an ideal threshold helps reduce the training time while simultaneously maintaining good performance. For example, select $t=20$$m$ as the ideal threshold for $K=50$, we could still achieve $95\%$ of the optimal performance but reduces required training time by roughly $20\%$. 

\begin{table}[htbp]
\centering
\caption{Ideal threshold}
\begin{tabular}{|c|c|c|c|}
\hline & $K=20$ & $K=50$ & $K=100$  \\
\hline Threshold ($m$) & $ \approx 30 $  & $ \approx 20 $ & $ \approx 15 $ \\
\hline
\end{tabular}
\label{tab:optimal_threshold}
\end{table}

\subsection{Time Complexity}
As shown in Section~\ref{subsection:time_complexity}, the time complexity hinges on the number of edges in the graph. we use the convention that capital letters denote the random variables and small letters for their realizations. 
Since the number of edges decreases as the threshold decreases, the time complexity inevitably depends on the threshold. 
Since the transceiver pair are close to each other, to approximately analyse such a relation, we use the median point $M_i$ between them to represent $i$-th transceiver pair. We denote the squared distance between the $i$-th and $j$-th transceiver pair by $Z_{i, j}=\left(X_{i}-X_{j}\right)^{2}+\left(Y_{i}-Y_{j}\right)^{2}$, where $X_i, Y_i$ and $X_j, Y_j$ are the coordinates of $M_i$ and $M_j$, respectively. We assume that $X_{i}, X_{j}, Y_{i}$ and $Y_{j}$ are i.i.d drawn from uniform distribution $\mathcal{U}(0,a)$, the cumulative distribution function of $Z_{i, j}$ for any two transceiver pairs are characterised in \eqref{eq:thresholdedges}.
We then define $N(t)$ as the number of edges left after the truncation with the threshold $t$. The expression for $N(t)$ is given by,
\begin{equation}
\begin{aligned}
N(t)&=\sum_{i=1}^{|\mathcal{V}|-1} \sum_{j=i+1}^{|\mathcal{V}|} \mathbbm{1}\left(Z_{i, j}<t^2\right),
\end{aligned}
\end{equation}
where $\mathbbm{1}$ is an indicator function and $|\mathcal{V}|$ is the total number of vertices. Here, we assume 
any two transceiver pairs have the same distance distribution. 
Since the probability $P\left(\mathbbm{1}\left(Z_{i, j}<t^2\right)=1\right)=P\left(Z_{i, j}<t^2\right) = F\left(t^{2}\right)$, the expectation is $ \mathbb{E}\left[\mathbbm{1}\left(Z_{i, j}<t^2\right)\right]=1 \cdot P\left(Z_{i, j}<t^2\right) = F\left(t^{2}\right)$. Since $Z_{i,j}$ has the same distribution for any two transceiver pairs, the expected number of edges left after the truncation with the threshold $t$ is given by
\begin{equation}
\begin{aligned}
\mathbb{E}[N(t)]  = \frac{|\mathcal{V}|(|\mathcal{V}|-1)}{2} \cdot F\left(t^{2}\right) = |\mathcal{E}|\cdot F\left(t^{2}\right).
\end{aligned}
\end{equation}
Since the time complexity for each layer is $\mathcal{O}(|\mathcal{V}|+|\mathcal{E}|)$ and it is dominated by $|\mathcal{E}|$,  the expected time complexity after truncation with the threshold $t$ will decrease by factor of $ F\left(t^{2}\right)$. Therefore, we could reduce the running time by introducing the threshold.

\section{Conclusion}

We proposed a novel graph representation for full-duplex transmission in power allocation problems and designed the corresponding graph neural network to find the optimal solution. The results showed that our algorithm has similar (and even slightly better) performance compared to traditional methods with less computation time. and strong generalisation capacities. Furthermore, we introduced an ideal threshold based on simulations to reduce time complexity and derived the analytic expression for expected time complexity for modelling such an effect. Further research needs to be conducted to be able to obtain the ideal threshold mathematically and generalise these results to different types of fading channels. 


\bibliographystyle{IEEEtran}
\bibliography{main}

\vspace{12pt}

\end{document}